# Coherent and Incoherent Coupling Dynamics between Neutral and Charged Excitons in Monolayer MoSe$_2$

Kai Hao[1], Lixiang Xu[1], Philipp Nagler[2], Akshay Singh[1], Kha Tran[1], Chandriker Kavir Dass[1], Christian Schüller[2], Tobias Korn[2], Xiaoqin Li[1,*], Galan Moody[3,*]

[1] Department of Physics and Center for Complex Quantum Systems, University of Texas at Austin, Austin, TX 78712, USA.

[2] Department of Physics, University of Regensburg, Regensburg, Germany 93040.

[3] National Institute of Standards & Technology, Boulder, CO 80305, USA.

## Abstract

The optical properties of semiconducting transition metal dichalcogenides are dominated by both neutral excitons (electron-hole pairs) and charged excitons (trions) that are stable even at room temperature. While trions directly influence charge transport properties in optoelectronic devices, excitons may be relevant through exciton-trion coupling and conversion phenomena. In this work, we reveal the coherent and incoherent nature of exciton-trion coupling and the relevant timescales in monolayer MoSe$_2$ using optical two-dimensional coherent spectroscopy. Coherent interaction between excitons and trions is definitively identified as quantum beating of cross-coupling peaks that persists for a few hundred femtoseconds. For longer times up to 10 ps, surprisingly, the relative intensity of the cross-coupling peaks increases, which is attributed to incoherent energy transfer likely due to phonon-assisted up-conversion and down-conversion processes that are efficient even at cryogenic temperature.

## Main Text

Quasiparticle dynamics in monolayer transition metal dichalcogenides (TMDs) have been the focus of intense experimental and theoretical research efforts due to the emergence of exotic spin-valley physics and growing interest in ultrathin electronics and optoelectronics.[1–6] Because of the heavy carrier effective masses and reduced dielectric screening, Coulomb interactions are at least an order of magnitude stronger in monolayer TMDs compared to conventional semiconductors as evidenced by neutral[7,8] and charged exciton (trion) states[9,10] with exceptionally large binding energies. The Coulomb interactions responsible for tightly bound states also enhance interactions between them. Evidence of coupling between excitons and trions has been observed in two-color pump-probe and photoluminescence spectroscopy experiments;[11,12] however, the coherent/incoherent nature of the interactions and the associated timescales remain unknown.

Because excitons and trions have different net charge, oscillator strength, and effective mass, coupling and transfer between them can substantially impact electron-hole recombination and charge transport for applications requiring high mobility and ultrafast optical modulation.[13,14] In the presence of coherent interactions between these quasiparticles, additional opportunities for quantum information and coherent control applications would emerge. Coherent coupling associated with non-radiative superpositions between states[15–17] has led to interesting quantum phenomena in conventional semiconductors including coherent population trapping, electromagnetically induced transparency, and lasing without inversion.[18–20] In the monolayer TMD $WSe_2$, up-conversion in photoluminescence has been reported, in which emission appears at the exciton resonance while resonantly pumping the trion.[12] Efficient energy up-conversion may find applications in laser cooling of solids.[21,22]

In this work, we probe the nonlinear optical response of exciton and trion transitions in monolayer $MoSe_2$ using optical two-dimensional coherent spectroscopy (2DCS), taking advantage of the technique's unique simultaneously high temporal and spectral resolutions.[23] We observe multiple coupling regimes on sub-picosecond and few-picosecond timescales corresponding to coherent and incoherent energy transfer, respectively.[24] The appearance of off-diagonal cross-coupling peaks in the 2D spectra are decisive signatures of exciton-trion interactions.[25] Oscillations in the cross-peak amplitudes at the exciton-trion difference frequency reveal that coupling is initially coherent, decaying with a 250 fs dephasing time. After dephasing of the quantum beats, enhancement in the relative strength of the cross peaks indicates remarkably efficient energy transfer via phonon-assisted exciton-to-trion down-conversion within 2-3 ps and trion-to-exciton up-conversion in ~10 ps. Such efficient energy transfer may be attributed to doubly resonant Raman scattering due to the fact that the exciton-trion splitting of ~30 meV is nearly resonant with the $A'_1$ optical phonon mode.[12,26]

We examine monolayer $MoSe_2$ mechanically exfoliated onto a sapphire substrate for optical transmission experiments. The sample is held in vacuum at a temperature of 20 K for the linear and nonlinear spectroscopy experiments. In $MoSe_2$, the lowest-energy exciton transition is between parallel electron and hole spin states in the upper and lower valence and conduction bands in the same momentum valley.[27,28] For the trion, the lowest energy configuration is the negative inter-valley singlet state formed between a spin-up electron-hole pair in the *K* valley and a single spin-down electron in the *K′* valley (Figure 1a). A time-integrated low-temperature photoluminescence spectrum obtained using 532 nm excitation is shown in Figure 1b (solid shaded region). The peaks at 1663 meV and 1632 meV are identified as the exciton (*X*) and trion (*T*). The trions are considered to be negatively charged due to unintentional *n*-type doping of the MoSe$_2$ crystal. The dashed curve depicts the excitation laser spectrum used for the nonlinear spectroscopy experiments, which is tuned to optically excite both transitions with similar fluence. For these experiments, the excitation laser spot size is ~30 μm full-width at half-maximum, which is smaller than the monolayer flake dimensions illustrated by the dashed outline in the white light optical image in Figure 1c.

The nonlinear optical response is characterized using 2DCS performed in the box geometry (Figure 1c). Optical 2DCS is a three-pulse transient four-wave mixing technique with the addition of interferometric stabilization of the pulse delays with femtosecond stepping resolution and nanosecond scan range. Details of the technique can be found in Ref. [29]. Briefly, three 40-fs pulses with wavevectors **k**$_1$, **k**$_2$, and **k**$_3$ interact with the sample to generate a four-wave mixing signal that is radiated in the phase-matched direction **k**$_S$ = -**k**$_1$ + **k**$_2$ + **k**$_3$. The experiments are performed in the rephasing time ordering, *i.e.* field $\mathcal{E}_1$ is incident on the sample first, and after interaction with fields $\mathcal{E}_2$ and $\mathcal{E}_3$ the resulting four-wave mixing signal field $S(t_1, t_2, t_3)$ is emitted as a photo echo for an inhomogeneously broadened system. The pulse time ordering is depicted in Figure 1d. Rephasing 2D spectra are generated by spectrally resolving the nonlinear signal through heterodyne interferometry with a fourth phase-stabilized reference pulse and scanning the delay $t_1$ between the first two pulses $\mathcal{E}_1$ and $\mathcal{E}_2$. Fourier transformation with respect to the delay $t_1$ produces a rephasing 2D spectrum of the signal $S(\hbar\omega_1, t_2, \hbar\omega_3)$ that correlates the excitation ($\hbar\omega_1$) and emission ($\hbar\omega_3$) energies of the system during the delays $t_1$ and $t_3$, respectively. The excitation fields and detected signal are co-circularly polarized for all experiments. The pump fluence is kept below 2 μJ/cm$^2$ (~9×10$^{11}$ excitons/cm$^2$), which is in the $\chi^{(3)}$ regime and well below saturation, in order to reduce contributions to the optical response from exciton-exciton dephasing and Auger recombination, which can broaden the transition linewidth and quench radiative recombination, respectively.[30–33]

A 2D amplitude spectrum is shown in Figure 2a for delay $t_2 = 0$ fs. The vertical and horizontal axes correspond to the excitation ($\hbar\omega_1$) and emission ($\hbar\omega_3$) energies, respectively. Quantum mechanical pathways (*e.g.* ground state bleaching, excited state emission, and non-radiative coherence) associated with each peak in the 2D spectrum are presented in the supporting information. The two peaks on the diagonal dashed line represent excitation and emission at the exciton (1665 meV, *X*) and trion (1634 meV, *T*) energies. The small ~2 meV Stokes shift of the resonances between the linear and nonlinear spectra indicates the excellent quality of the material. The peaks are elongated along the diagonal due to inhomogeneous broadening from impurity and defect potentials in the material, whereas broadening along the cross-diagonal is determined by the homogeneous dephasing rate $\gamma$ (interband optical coherence time $T_2 = \hbar/\gamma$) of each transition.[34] From fits to the lineshapes, we find for the exciton $\gamma_X$ = 1.4±0.2 meV ($T_2$ = 470±60 fs) and for the trion $\gamma_T$ = 1.3±0.2 meV ($T_2$ = 510±70 fs) (see supporting information).

The signature of exciton-trion coupling is the appearance of a lower (higher) off-diagonal cross-coupling peak *LCP* (*HCP*), which originates from excitation at the exciton (trion) energy and emission at the trion (exciton) energy. In principle, the cross-coupling peaks can appear from both coherent coupling and incoherent energy transfer between states. These processes arise from different microscopic effects: the former comes from the light-matter interaction with the first two fields $\mathcal{E}_1$ and $\mathcal{E}_2$ that drives the system into a Raman-like non-radiative coherent superposition between the excited exciton and trion states, which oscillates during the delay $t_2$ at their difference frequency.[35,36] Raman coherence beats have been observed previously in semiconductor bulk, quantum wells, and quantum dots.[37–45] Conversely, the latter arises from incoherent energy transfer between states due to phonon-assisted up-conversion (*HCP*) and down-conversion (*LCP*) processes.

To disentangle coherent and incoherent coupling mechanisms, we further acquired 2D spectra for increasing delay $t_2$, shown in Figure 2b-d for 70 fs, 140 fs, and 6 ps. Each spectrum is normalized to the maximum value of the exciton resonance (relative scales are indicated above each panel). The amplitudes of the cross-coupling peaks oscillate nearly in phase during $t_2$, which is direct evidence of coherent exciton-trion coupling. To support this interpretation, we model the coherent nonlinear response using a perturbative expansion of the density matrix up to third order in the excitation field for a four-level 'diamond' system, illustrated in Figure 3a (see supporting information for details). Interactions are introduced phenomenologically[46] by breaking the symmetry between the upper and lower transitions through a shift of state $|XT\rangle$. Simulated spectra are shown in Figure 3a for increasing delay $t_2$, which clearly show the quantum beating of the cross-coupling peaks.

A comparison between the measured and simulated *LCP* and *HCP* amplitudes is shown in Figure 3b by the symbols and solid curves, respectively. The measured and simulated peak amplitudes are averaged along $\hbar\omega_1$ and $\hbar\omega_3$ within a ±15 meV window of each peak, which enhances the signal-to-noise ratio and minimizes contributions from spectral diffusion processes. Interference between oscillating Raman-like coherence terms and exponentially decaying phase-space filling nonlinearities reduce the visibility of the quantum beat amplitude from unity. From the simulations we extract the beat period ($\tau_{XT}$) and the dephasing time of the exciton-trion coherence ($\tau_c$). The quantum beat period $\tau_{XT} = 132\pm20$ fs corresponds to an exciton-trion splitting of $\Delta_{XT}= 2\pi\hbar/\tau_{XT} \approx 31\pm4$ meV, which is in excellent agreement with the trion binding energy from the linear and nonlinear spectra. We find that the quantum beat dephasing time is $\tau_c = 250\pm30$ fs ($\gamma_{XT} = \hbar/\tau_c = 2.6\pm0.2$ meV). The fact that $\gamma_{XT}$ is equivalent to the sum of the exciton ($\gamma_X = $ 1.4 meV) and trion ($\gamma_T = $ 1.3 meV) dephasing rates implies that dissipative fluctuations that broaden the exciton and trion transition linewidths are uncorrelated,[47] which is in contrast to anti-correlated heavy-hole—light-hole exciton dephasing in GaAs quantum wells[48] and correlated dephasing of fine-structure exciton states in InAs quantum dots.[49] For the *LCP*, the non-oscillating component arises from ground-state bleaching and excited state absorption of the exciton transition, which exhibits a faster recombination lifetime compared to the trion (see Figure 4 and supporting online information). As a result, the non-oscillating component of *LCP* decays faster compared to *HCP*, which is associated with the longer-lived trion population.

The amplitude of the cross-coupling peaks is surprisingly large for spectra taken at $t_2$ longer than a few picoseconds (*e.g.* the spectrum for $t_2 = 6$ ps in Figure 2d). We attribute the enhanced *LCP* and *HCP* amplitudes relative to the diagonal *X* and *T* peaks to phonon-assisted exciton-to-trion down-conversion and trion-to-exciton up-conversion processes, respectively. The exciton and trion conversion efficiency is enhanced in monolayer MoSe$_2$ owing to a doubly resonant Raman scattering process involving a single optical $A'_1$ phonon, which has energy similar to the ~30 meV trion binding energy.[12,26,50] The up-conversion (*HCP*) and down-conversion (*LCP*) dynamics are shown in Figure 4 for delays $t_2$ up to 25 ps. The peaks are normalized using the procedure described in the supporting information. To quantify the incoherent energy transfer times, we use a rate equation analysis that takes into account exciton and trion interband recombination, bi-directional scattering between bright and dark states, and bi-directional exciton-trion energy transfer (see supporting information for details). Fully constrained results are obtained by simultaneously fitting *X*, *T*, *LCP*, and *HCP*, which are shown by the solid lines in Figure 4. We obtain an $X \rightarrow T$ conversion time of 2.5±0.2 ps, which is comparable to a previous pump-probe study on trion formation.[51] Conversely, the time required for $T \rightarrow X$ up-conversion is expected to be longer since this is an anti-Stokes scattering process involving trion dissociation into an exciton and free electron accompanied by phonon absorption to conserve energy and momentum, which depends on the average phonon occupation given by a Bose-Einstein distribution. Consistent with this notion, we find the $T \rightarrow X$ conversion time is 8±1 ps. Both conversion processes are a factor of 3-5 faster compared to calculations for delocalized excitons in a WSe$_2$ monolayer.[32] This difference might be explained by the moderate exciton and trion localization due to impurity and defect potentials observed here, which is also responsible for the inhomogeneous broadening in this sample. With momentum no longer being a good quantum number, conservation constraints are relaxed and the efficiency of the conversion processes is enhanced.

The amplitudes of the diagonal peaks *X* and *T* associated with exciton and trion population relaxation are also analyzed with the rate equation model. Interestingly, fits using this model indicate that compared to interband recombination, faster bright-to-dark state scattering is required in order to reproduce the biexponential population decay dynamics of *X* and *T*. The long-lived components are attributed to repopulation of the bright states from the dark states, which might be associated with states outside of the light cone with large momentum.

In conclusion, we provide a comprehensive picture of quasiparticle coupling, energy transfer, and relaxation dynamics in monolayer MoSe$_2$, enabled by the unique capability of 2DCS in tracking ultrafast dynamics of multiple resonances and disentangling different quantum mechanical pathways. Oscillations of the cross-coupling peaks in the two-dimensional spectra provide unambiguous evidence of coherent exciton-trion interactions, which persist for a few hundred femtoseconds. Robust quantum coherence in TMDs may lead to engineerable wave-like energy transport between delocalized quantum states, similar to concepts in photosynthesis, with the potential to enhance future photovoltaic efficiency.[52,53] On a longer timescale up to 10 ps, relative enhancement of these cross peaks is a signature of phonon-assisted up-conversion and down-conversion processes that are surprisingly efficient even at cryogenic temperatures. In addition to influencing the exciton and trion recombination dynamics, coherent and incoherent coupling may be leveraged for the generation and control of long-lived valley coherence through resonant transfer of the valley information to the electronic spin state of the trion.[54]

## Author Information

**Corresponding Authors**

*E-mail (G.M.): galan.moody@nist.gov

*E-mail (X.L.): elaineli@physics.utexas.edu

The authors declare no competing financial interest.

## Acknowledgments

The authors gratefully acknowledge fruitful discussion with R. Huber. The work performed by K. Hao. K, Tran, A. Singh, and X. Li were supported jointly by NSF DMR-1306878 and NSF EFMA-1542747. The work performed by L. Xu and X. Li was supported as part of the SHINES, an Energy Frontier Research Center funded by the U.S. Department of Energy (DOE), Office of Science, Basic Energy Science (BES) under Award # DE-SC0012670. X. Li also acknowledges the support from a Humboldt fellowship. P. Nagler, C. Schüller and T. Korn gratefully acknowledge technical assistance by S. Bange and financial support by the German Research foundation (DFG) via GRK 1570 and KO3612/1-1.

## Figure Captions

**Figure 1:** (a) Band diagram illustration of the lowest energy trion (left panel) and exciton (right panel) states in monolayer $MoSe_2$. The exciton and trion are coupled coherently through Coulomb interactions and incoherently phonon- or defect-assisted energy transfer. (b) Photoluminescence spectrum taken at 20 K. The peaks at 1663 meV and 1632 meV are attributed to the exciton ($X$) and trion ($T$) resonances, respectively. The laser spectrum for the nonlinear experiments is depicted by the dashed curve. (c) Schematic diagram of the box-geometry used for the three-pulse four-wave mixing experiments. The left panel shows an optical image of monolayer $MoSe_2$ on sapphire (within the dashed outline) and the laser spot for photoluminescence experiments. (d) The pulse time ordering for the nonlinear spectroscopy experiments.

**Figure 2**: Rephasing 2D coherent spectra acquired for various $t_2$ delays and co-circular polarization of the excitation pulses and detected signal. The spectra correlate the excitation ($\hbar\omega_1$) and emission ($\hbar\omega_3$) energies of the system. The exciton ($X$) and trion ($T$) peaks appear on the diagonal dashed line, whereas the higher (*HCP*) and lower (*LCP*) cross-coupling peaks that oscillate with increasing $t_2$ are clear signatures of coherent exciton-trion coupling (a-c). At long timescales ($t_2 > 1$ ps), $X \rightarrow T$ and $T \rightarrow X$ incoherent energy transfer appear at the *LCP* and *HCP* energies, respectively (d).

**Figure 3:** (a) 2D spectra simulated using a perturbative expansion of the density matrix for a four-level diamond system. (b) Quantum beats of the *LCP* and *HCP* amplitudes versus delay $\boldsymbol{t_2}$ on a sub-picosecond timescale (symbols). Agreement between the measurements and the model (solid lines) provides the dephasing time ($\tau_c \approx 250$ fs) and period ($\tau_{XT} \approx 130$ fs) of the quantum beats.

**Figure 4:** Amplitudes of the exciton (*X*), trion (*T*), lower (*LCP*), and higher (*HCP*) cross-coupling peaks versus delay $t_2$ on a picosecond timescale. The data are modeled with the rate equations discussed in the supporting information (solid lines).

**Figure 1**

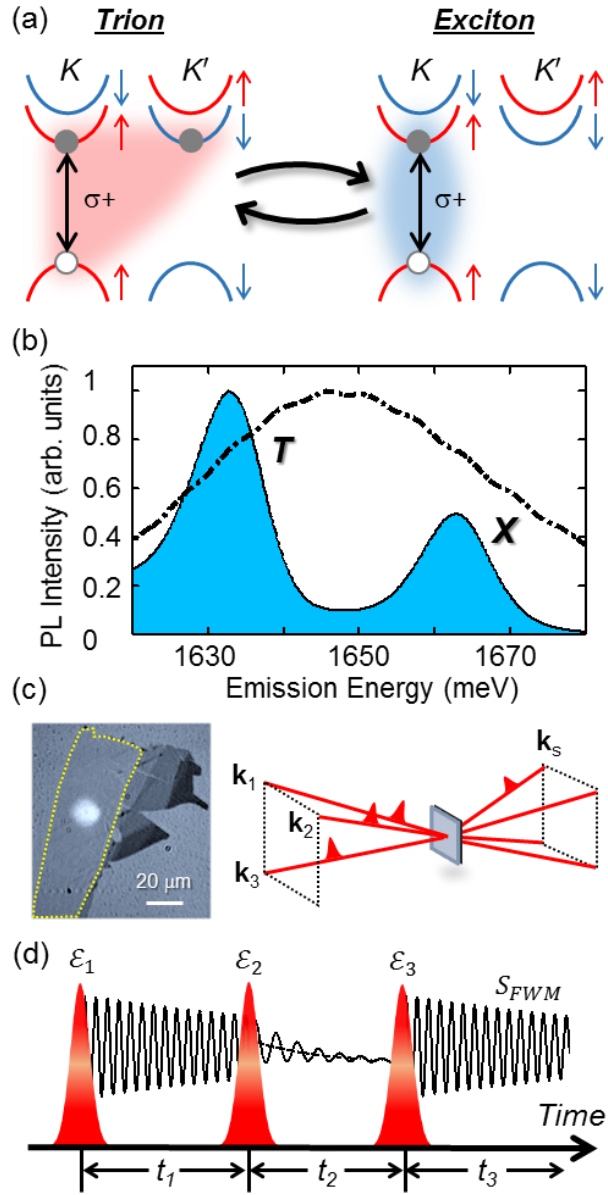

**Figure 2**

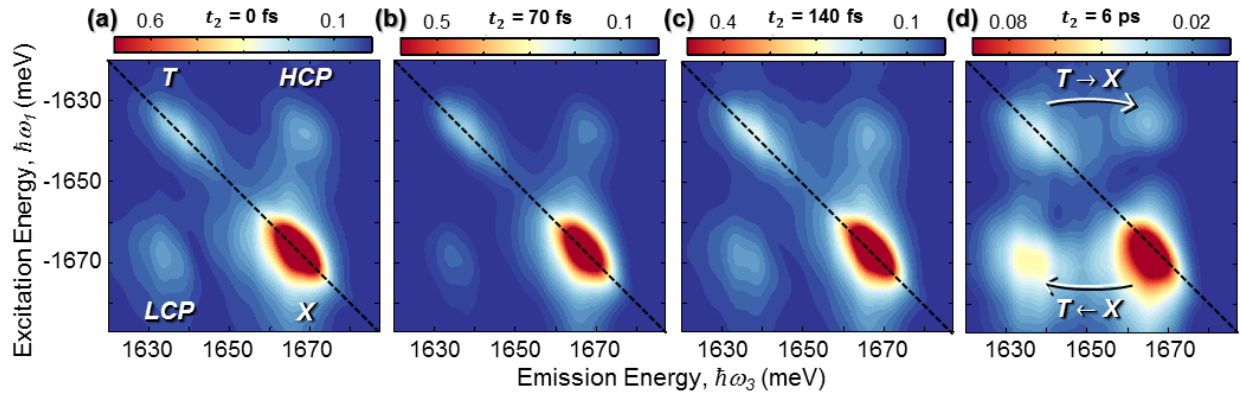

**Figure 3**

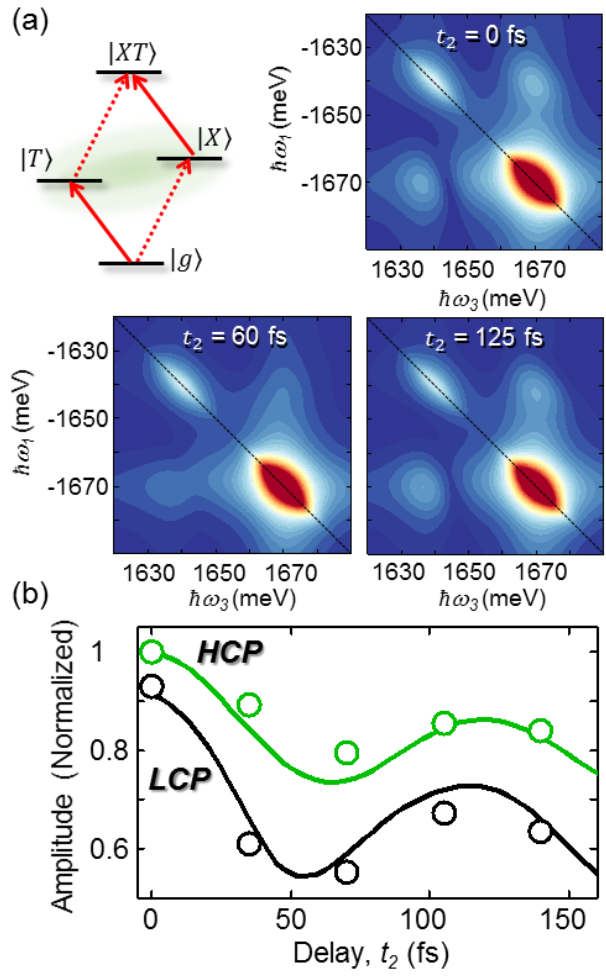

**Figure 4**

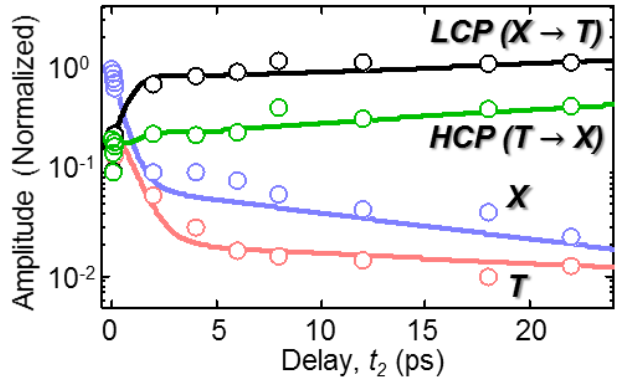

# Supporting Information for: Coherent and Incoherent Coupling Dynamics between Neutral and Charged Excitons in Monolayer MoSe$_2$

Kai Hao[1], Lixiang Xu[1], Philipp Nagler[2], Akshay Singh[1], Kha Tran[1], Chandriker Kavir Dass[1], Christian Schüller[2], Tobias Korn[2], Xiaoqin Li[1,*], Galan Moody[3,*]

[1] Department of Physics and Center for Complex Quantum Systems, University of Texas at Austin, Austin, TX 78712, USA.

[2] Department of Physics, University of Regensburg, Regensburg, Germany 93040.

[3] National Institute of Standards & Technology, Boulder, CO 80305, USA.

## S1. Exciton Excitation Density

Linear absorption measurements are performed by recording white light differential transmission through the sample and substrate, which is shown in Figure S1 below. The linear absorption allows for the exciton excitation density to be determined from the following expression:

$$N_X = \frac{P_{ave}T_p(1-R)(1-e^{-\alpha L})}{\pi r^2 E_{ph}}, \qquad (S1)$$

where $P_{ave}$ is the average power per beam, $T_p$ = 12.5 ns is the laser pulse period, $R = 0.15$ takes into account reflection losses, $A = 1 - e^{-\alpha L} = 0.15$ is the peak linear absorbance of the MoSe$_2$ monolayer, $r$ = 17.5 μm is the focused beam radius, and $E_{ph}$ = 1665 meV is the average photon energy. We note that only the maximum value of the linear absorbance at ~1665 meV is used to determine the excitation density, which overestimates the actual density in the experiments.

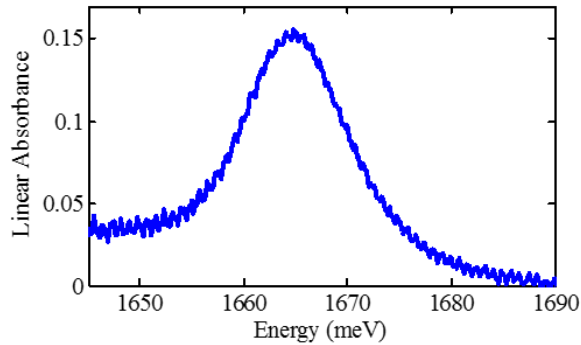

**Figure S3.** Linear absorbance at the exciton resonance.

## S2. Exciton and Trion Homogeneous Linewidths

As an approximation, homogeneous and inhomogeneous broadening appear in a rephasing two-dimensional amplitude spectrum along the cross-diagonal and diagonal directions of a diagonal peak, respectively. In the case of moderate inhomogeneity where the homogeneous (γ) and inhomogeneous (σ) linewidths are comparable, these contributions become entwined in the spectrum. Following Ref. (1), to extract the homogeneous linewidth we simultaneously fit cross-diagonal ($A_{CD}$) and diagonal ($A_D$) lineshapes to the expressions below:

$$A_{CD} = \frac{e^{\frac{(\gamma - i\hbar\omega)^2}{2\sigma^2}} Erfc\left(\frac{\gamma - i\hbar\omega}{\sqrt{2}\sigma}\right)}{\sigma(\gamma - i\hbar\omega)}$$

$$A_D = \frac{\sqrt{2\pi}}{\gamma} Voigt(\gamma, \sigma, \hbar\omega), \qquad (S2)$$

where $\hbar\omega$ is the emission energy, *Erfc* is the complementary error function, and the *Voigt* profile is a convolution of Gaussian and Lorentzian functions. Homogeneous and inhomogeneous lineshapes are shown in the top and bottom panels, respectively, of Figure S2 taken at the peak of the exciton (left) and trion (right) transitions. The lineshapes are simultaneously fit with Eqn. S2 (solid lines). The exciton and trion homogeneous linewidths extracted from the fits are $\gamma_X$ = 1.4 meV ($T_2 = \hbar/\gamma$ = 470 fs) and $\gamma_T$ = 1.3 meV ($T_2$ = 510 fs). The inhomogeneous linewidths are $\sigma_X$ = 3.6 meV and $\sigma_T$ = 4.8 meV (corresponding to a Gaussian full-width at half-maximum of $2\sqrt{2ln2}\,\sigma$).

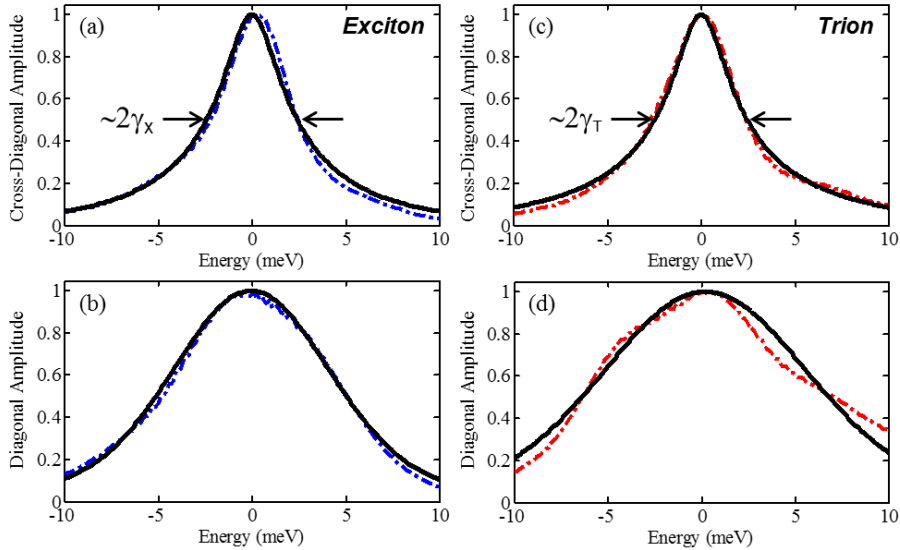

**Figure S4.** The exciton (a) cross-diagonal and (b) diagonal lineshapes simultaneously fit using Eqn. (S2). The half-width at half-maximum of the cross-diagonal fit is approximately equal to the exciton homogeneous linewidth $\gamma_X$. Cross-diagonal and diagonal lineshapes and fits for the trion are shown in (c) and (d), respectively.

For a two-level system, the optical coherence time can be expressed in terms of the population recombination lifetime ($T_1$) and the elastic pure dephasing time ($T_2^*$) according to

$$\frac{1}{T_2} = \frac{1}{2T_1} + \frac{1}{T_2^*}. \tag{S3}$$

Interesting, we find that for the exciton, $T_2 \approx 2T_1$, indicating lifetime-limited dephasing. On the other hand, for the trion we find that $T_2^* = 700$ fs, revealing that under the same experimental conditions, the trion coherence decays from population relaxation and pure dephasing on comparable timescales. These results are summarized in Table S1 below. The uncertainties are estimated from the value at which the least-squares error between the measurements and the fits changes by 50%.

**Table S1.** Exciton and trion optical dephasing and relaxation times.

|  | $\gamma\ (T_2)$ | $\sigma$ | $\Gamma\ (T_1)$ | $\gamma^*(T_2^*)$ |
|---|---|---|---|---|
| ***Exciton*** | 1.4±0.2 meV (470±60 fs) | 3.6±0.3 meV | 2.3±0.2 meV (290±30 fs) | 0.25±0.2 meV (3±1 ps) |
| ***Trion*** | 1.3±0.2 meV (510±70 fs) | 4.8±0.4 meV | 0.66±0.1 meV (1±0.1 ps) | 1±0.3 meV (700±200 fs) |

## S3. Perturbative Density Matrix Calculations

We model exciton and trion coherent coupling using a four-level diamond system as shown in Figure S3. We can represent two independent two-level systems [Figure S3a] as a four-level system through a Hilbert space transformation [Figure S3b]. The two representations are equivalent provided the lower transitions ($|g'\rangle \leftrightarrow |T'\rangle$ and $|g'\rangle \leftrightarrow |X'\rangle$) have the same properties (transition energy, dipole moment, dephasing rate) as the corresponding upper transitions ($|X'\rangle \leftrightarrow |XT'\rangle$ and $|T'\rangle \leftrightarrow |XT'\rangle$). We note that the states in Figure S3b are essentially two-particle states, *e.g.* state $|T'\rangle$ is the composite of the trion in the excited state and the exciton in the ground state. We limit the number of levels to the doubly excited state since only quantum pathways up to this level can contribute to the four-wave mixing signal in the $\chi^{(3)}$ regime.

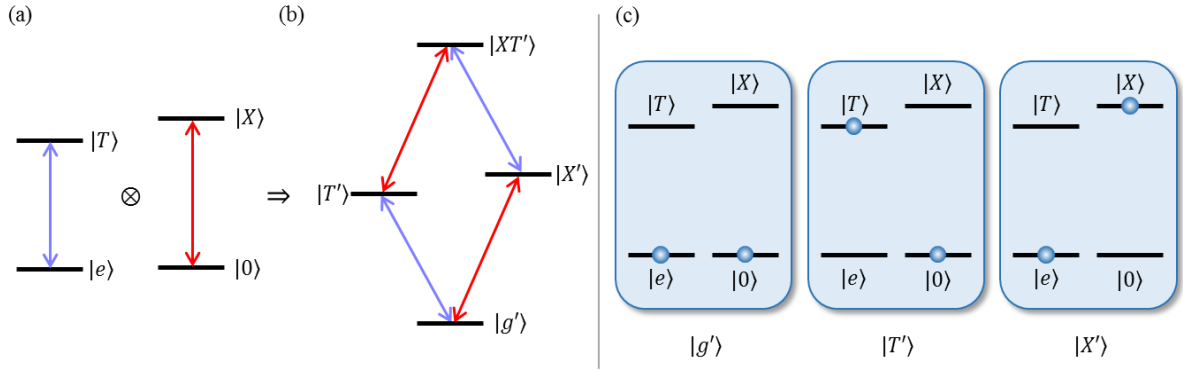

**Figure S5.** Through a Hilbert space transformation, two independent two-level systems (a) can be represented as a single four-level diamond system (b). The resulting singly excited exciton ($|X'\rangle$) and trion ($|T'\rangle$) states are represented in (c). Exciton-trion interactions are phenomenologically introduced by breaking the symmetry between the upper and lower transitions in (b).

The nonlinear response is modeled by perturbatively solving the density matrix up to third-order in the excitation field. Details of this analysis can be found in Ref. (2). The resulting third-order coherent response for the four-level system can be represented by the double-sided Feynman diagrams shown in Figure S4. In total, twelve diagrams contribute to the signal—two each for the exciton and trion, and four each for the lower and higher cross-coupling peaks. The pathways for the exciton and trion arise from excited-state emission [(1) and (3)] and ground-state bleaching [(2) and (4)] nonlinearities. For the cross peaks, ground-state bleaching, excited-state emission, and excited-state absorption nonlinearities contribute; however, in the absence of interaction effects, the signals associated with the quantum pathways involving the upper transitions [(6) and (8) for *LCP*, (10) and (12) for the *HCP*] destructively interfere with the signals involving just the lower transitions [(5) and (7) for *LCP*, (9) and (11) for the *HCP*]. Therefore, no cross-coupling peaks will appear in the 2D spectra.

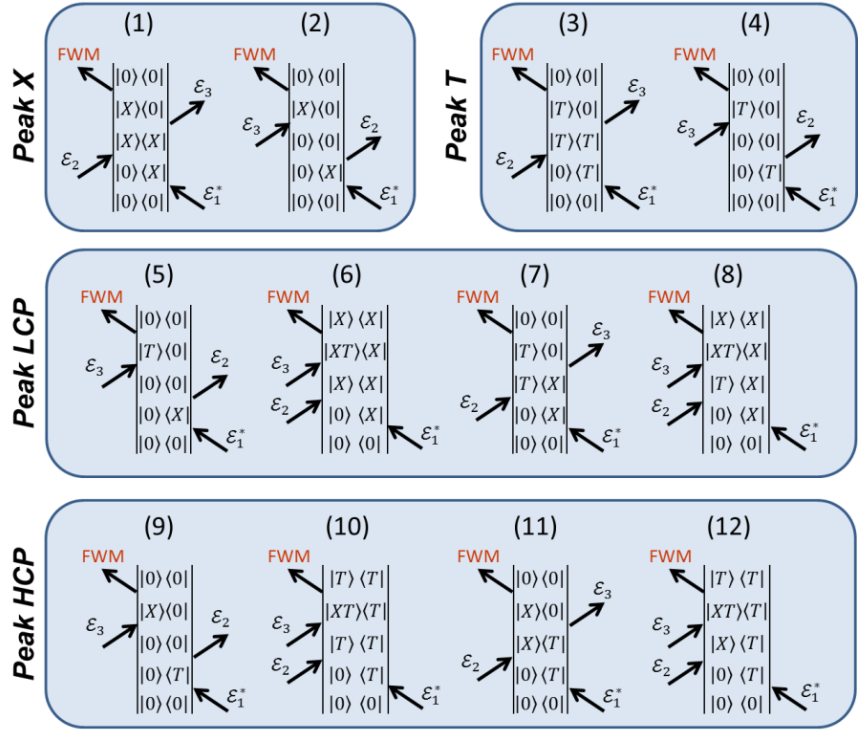

**Figure S6.** Quantum pathways represented by Feynman diagrams used for the density matrix calculations.

Interactions giving rise to the cross-coupling peaks can be phenomenologically introduced by breaking the equivalence of the upper and lower transitions through modifications to the excited state transition energies, dephasing rates, or dipole moments. To model the measured spectra in the main text, we introduce a one-meV energy shift of the excited state, which breaks the symmetry between diagrams (5) - (8) for the *LCP* and diagrams (9) - (12) for the *HCP*. The homogeneous and inhomogeneous linewidths are taken from the cross-diagonal and diagonal lineshapes discussed above. The lifetimes are obtained from the measured amplitude dynamics versus delay $t_2$ shown in Figure 4 in the main text. The exciton and trion transition energies are obtained from the peak positions in the 2D spectra. The ratio of the exciton and trion transition dipole moments is adjusted to match the amplitudes of peaks *X* and *T* at zero delay. The remaining free parameter is the dephasing time of the exciton-trion coherence, which is used as a fit parameter to the data discussed in the main text. The quoted uncertainties correspond to the value at which the least-squares error between the measurements and the model increases by 50%.

## S4. Rate Equation Analysis Parameters.

We use a series of coupled rate equations for a three-level system describing each of the bright and dark exciton and trion resonances, as illustrated in Figure S5. We consider multiple relaxation channels including population relaxation ($\Gamma_B$), bi-directional scattering of bright states to long-lived dark states ($\Gamma_{BD}$ and $\Gamma_{DB}$), and bi-directional exciton-trion energy transfer ($\Gamma_{XT}$ and $\Gamma_{TX}$). These processes are captured by the following expressions:

$$\frac{dN_B^X}{dt} = -[\Gamma_B^X + \Gamma_{BD}^X + \Gamma_{XT}]N_B^X + \Gamma_{TX}N_B^T + \Gamma_{DB}^X N_D^X$$

$$\frac{dN_B^T}{dt} = -[\Gamma_B^T + \Gamma_{BD}^T + \Gamma_{TX}]N_B^T + \Gamma_{XT}N_B^X + \Gamma_{DB}^T N_D^T$$

$$\frac{dN_D^X}{dt} = -\Gamma_{DB}^X N_D^X + \Gamma_{BD}^X N_B^X$$

$$\frac{dN_D^T}{dt} = -\Gamma_{DB}^T N_D^T + \Gamma_{BD}^T N_B^T$$

where $N_B^i$ and $N_D^i$ correspond to the optically bright and dark state populations for transition $i = X$, $T$. The $X$ and $T$ amplitudes are normalized to the maximum exciton amplitude at zero delay. The rate equations are solved with and without the exciton-trion coupling terms ($\Gamma_{TX}N_B^T$ and $\Gamma_{XT}N_B^X$) to determine the *LCP* and *HCP* amplitudes. The amplitudes of *HCP* and *LCP* are normalized to the geometric mean of $X$ and $T$, which compensates for any differences in oscillator strength and spectral overlap of the excitation pulses with each resonance. The rate equations are simultaneously fit to all amplitudes and the results are shown by the solid curves in Figure 4 of the main text. The quoted uncertainties correspond to the value at which the least-squares error between the measurements and the model increases by 50%.

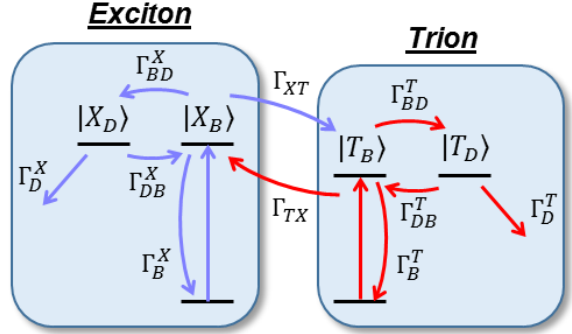

**Figure S7.** The rate equations are modeled with the above energy level diagrams for the exciton and trion.

Quantitative agreement between the model and measurements of all peaks provides insight into the dominant relaxation and coupling channels. The ultrafast population relaxation component measured for the exciton (~300 fs) and the trion (~1 ps) is ascribed to a competition between the bright state relaxation channels shown in Figure S5. Following resonant optical excitation, we consider the creation of excitons and trions at the band edge with negligible center-of-mass momentum. From fits of the model to the data, we find that the fastest relaxation channel for both the exciton and trion is to the dark state with rates $\Gamma_{BD}^X = 1.1$ meV (600 fs) and $\Gamma_{BD}^T = 0.6$ meV (1.1 ps), respectively. The dark states $|X_D\rangle$ and $|T_D\rangle$ might be associated with intra-valley scattering of bright excitons and trions to dark states outside of the light cone with momentum larger than $n\omega/c$. Redistribution of excitons in momentum space through exciton-exciton and exciton-impurity intra-valley scattering ($\Gamma_{BD}$) can be efficient due to weak band dispersion near the *K/K'* points resulting from the heavy electron and hole effective masses. Repopulation of the bright states from the dark states leads to the biexponential decay dynamics observed for peaks *X* and *T*. We find that the relative weight of the fast and slow decay dynamics of peaks *X* and *T* is most sensitive to the bi-directional bright-dark state scattering rates ($\Gamma_{BD}$ and $\Gamma_{DB}$) and the interband recombination rates ($\Gamma_B$). The former primarily affects the fast decay component lifetime and the relative amplitude of the fast and slow components, whereas the latter primarily determines the lifetime of the slow component. From the fits we find that $\Gamma_B^X = 0.3$ meV (2.2 ps) and $\Gamma_B^T = 0.26$ meV (2.5 ps). In principle this population relaxation channel can be ascribed to both radiative and non-radiative decay mechanisms. For the exciton, bright-to-dark state scattering ($\Gamma_{BD}^X$) occurs within ~600 fs, which, in combination with a picosecond interband recombination time ($\Gamma_B$) and few-picosecond exciton-trion down-conversion ($\Gamma_{XT}$), leads to the fast decay time of peak *X* (~300 fs). The dark exciton states repopulate the near-zero center-of-mass bright states with an intra-valley relaxation rate $\Gamma_{DB}^{X} = 0.13$ meV (5 ps), feeding the slow decay dynamics of peak *X*. For the trion, intra-valley relaxation occurs at a rate $\Gamma_{DB}^{T} = 0.04$ meV (18 ps), which is slower compared to the exciton possibly due to the larger effective mass of the trion.[3] The parameters used for the fits to the data are given in Table S2 below.

**Table S3.** Parameters for the rate equation analysis of the amplitudes in Figure 4b of the main text.

| [meV] | $\Gamma_B$ | $\Gamma_D$ | $\Gamma_{BD}$ | $\Gamma_{DB}$ | $\Gamma_{XT}$ | $\Gamma_{TX}$ |
|---|---|---|---|---|---|---|
| **Exciton** | 0.3±0.04 (2.2±0.1 ps) | $10^{-3}$ (0.7 ns) | 1.1±0.2 (0.6±0.07 ps) | 0.13±0.02 (5±0.9 ps) | 0.27±0.02 (2.5±0.2 ps) | - |
| **Trion** | 0.26±0.09 (2.5±0.3 ps) | $10^{-3}$ (0.7 ns) | 0.6±0.04 (1.1±0.1 ps) | 0.04±0.01 (18±2 ps) | - | 0.08±0.01 (8±1 ps) |